\documentclass[3p]{elsarticle}

\usepackage{color}
\usepackage{amssymb}
\usepackage{natbib}
\usepackage{longtable}
\usepackage{tabulary}
\biboptions{numbers,sort&compress}
\usepackage{caption2}
\usepackage{amssymb}
\usepackage{graphicx}
\usepackage[colorlinks,citecolor=blue,linkcolor=blue,urlcolor=black,hyperindex,CJKbookmarks]{hyperref}
\usepackage{url}
\usepackage{booktabs}
\usepackage{colortbl}
\usepackage{makecell}
\usepackage{enumerate}
\usepackage{algorithm}
\usepackage{algorithmic}
\usepackage{soul}

\usepackage{color} 
\definecolor{mypink2}{RGB}{219, 48, 122}
\definecolor{orange}{RGB}{255, 147, 00}
\definecolor{grey}{RGB}{166, 166, 166}

\usepackage{amsmath}
\usepackage{amsthm}
\graphicspath{{picture/}}

\theoremstyle{definition}

\allowdisplaybreaks

\newcommand{\bea}{\begin{eqnarray}}
	\newcommand{\eea}{\end{eqnarray}}
\newcommand{\ben}{\begin{equation}}
	\newcommand{\een}{\end{equation}}

\makeatletter
\let\@footnotetext\@gobble
\makeatother

\begin{document}

\begin{frontmatter}

\title{EMIF: Evidence-aware Multi-source Information Fusion Network for Explainable Fake News Detection}

\author[dut1]{Qingxing Dong\corref{cor1}}
\cortext[cor1]{Corresponding author.}
\ead{qxdong@whu.edu.cn,qingxingdong@gmail.com}
\author[dut2]{Mengyi Zhang}
\ead{mengyi\_zhang@outlook.com}
\author[dut3]{Shiyuan Wu}
\ead{shiyuanwu@mails.ccnu.edu.cn}
\author[dut1]{Xiaozhen Wu}
\ead{wxzher@outlook.com}

\address[dut1]{School of Journalism and Communication, Wuhan University, Wuhan 430072, China}
\address[dut2]{National Institute of Cultural Development, Wuhan University, Wuhan 430072, China}
\address[dut3]{School of Information Management, Central China Normal University, Wuhan 430072, China}

\begin{abstract}
In responcse to the significant detrimental effects of fake news proliferation, extensive research on automatic fake news detection has been conducted. Most existing approaches rely on a single source of evidence, such as comments or relevant news, to derive explanatory evidence for decision-making, demonstrating exceptional performance. However, their single evidence source suffers from two critical drawbacks: (i) noise abundance, and (ii) resilience deficiency. Inspired by the natural process of fake news identification, we propose an Evidence-aware Multi-source Information Fusion (EMIF) network that jointly leverages user comments and relevant news to make precise decision and excavate reliable evidence. To accomplish this, we initially construct a co-attention network to capture general semantic conflicts between comments and original news. Meanwhile, a divergence selection module is employed to identify the top-K relevant news articles with content that deviates the most from the original news, which ensures the acquisition of multiple evidence with higher objectivity. Finally, we utilize an inconsistency loss function within the evidence fusion layer to strengthen the consistency of two types of evidence, both negating the authenticity of the same news. Extensive experiments and ablation studies on real-world dataset FibVID show the effectiveness of our proposed model. Notably, EMIF shows remarkable robustness even in scenarios where a particular source of information is inadequate.
\end{abstract}


\end{frontmatter}


\section{Introduction}
Compared to traditional information carriers like newspapers and magazines, social media platforms offer instant access to massive information from various sources, including official channels, social accounts, citizen journalists, and interactive spaces like comment sections. However, due to the convenience of disseminating false information and lack of platform-supervision, the proliferation of fake news on social media has reached alarming proportions. The widespread dissemination of fake news may have far-reaching consequences, including sowing chaos, inciting hatred, eroding trust, and infiltrating various aspects of individual lives, politics, economics, and societal harmony \cite{Wasserman2019}. For instance, the massive infodemic during COVID-19 negatively impacted people’s mental or physical well-being and strained the public healthcare systems \cite{Islam2020}. Therefore, it becomes an essential urgency to develop automatic fake news detection systems, which can further contribute to the purification and harmonization of the online information ecosystem.

Initially, researchers turned to traditional machine learning methods like Support Vector Machine (SVM) \cite{Ahmed2017} as baseline models for fake news detection. However, their heavy reliance on features engineering introduced inevitable subjective bias. In recent years, deep learning based models, such as Recurrent Neural Networks (RNNs) \cite{Ma2016} and Convolutional Neural Networks (CNNs) \cite{Yu2017}, have gained great prominence in this field. Yet, these models often act as black boxes, diminishing their trustworthiness and practical utility. While the development of fully transparent white-box models offers a potential solution, it may come at the expense of predictive performance and is still in its early stages. Alternatively, post-hoc interpretation techniques, like feature correlation methods \cite{Ayoub2021}, offer a means to elucidate the contribution of different features. This provides decision-makers with valuable insights for interpreting model results. However, these models often fail to convert the explanations derived from the perspective of models into human-centered designs, posing a challenge for operators in fact-checking agencies to rely on understandable evidence for decision-making. As advocated by the academic community, we should build an Explainable AI (XAI) system that encompasses both interpretable models and explanatory evidence \cite{Arrieta2020} \cite{Gunning2019}. Currently, significant efforts in fake news detection have devoted to excavating interactive evidence from related content sources, such as comments \cite{Shu2019a} \cite{Ge2022} and relevant news \cite{Wu2021b} \cite{Nie2019}, while analyzing attention weights \cite{Song2021}.

\begin{figure}[htbp] 
	\centering 
	\includegraphics[width = 6in]{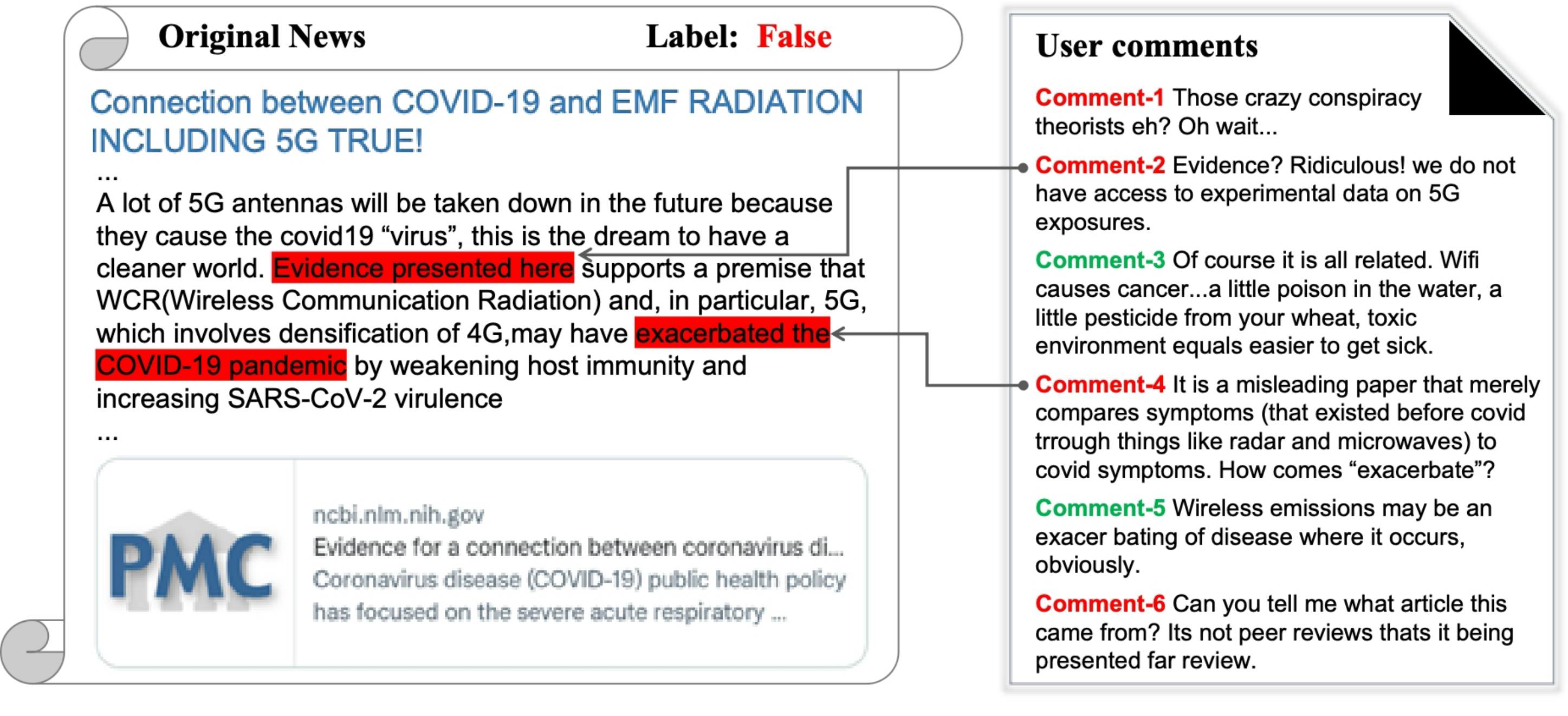} 
	\caption{A piece of news and its related comments on social media. The green and red comments express supporting and opposing viewpoints, respectively. But only Comment-2 and Comment-4 point out the core false part of this fake news.} 
	\label{f1}
\end{figure}

Despite possessing commendable effectiveness, these evidence sources central to current methodologies exhibit noteworthy shortages. (i) \textbf{Noise abundance}. User comments, a primary input in fake news detection tasks, may not consistently reveal reflective and authentic views from users due to various intractable objective biases \cite{Wang2023} and intentional opinion manipulation. Meanwhile, not all user comments effectively address false elements in news articles, leading to semantic conflicts unsuitable for supporting detection results, as illustrated in Fig. \ref{f1}. Furthermore, recent advances in large language models (LLMs) have heightened capabilities for generating convincing and nuanced related news at an unprecedented scale \cite{Pan2023}. The failure to filter out excess noises in these evidence perpetuates the trade-off dilemma between interpretability and performance. (ii) \textbf{Resilience deficiency}. Most existing methods rely on a single information source as their model input, which can bring about additional risks and costs. For example, user comments frequently contain intentional deletions and accounts hijacking \cite{Sun2022}, while relevant news may be limited in availability, damaged, or belated accessibility \cite{Wu2021a}. This results in a lack of model robustness and hampers its ability to generalize to more adversarial scenarios.

To address the mentioned issues, we draw inspiration from how people naturally recognize fake news by focusing on critical semantic conflicts within user comments and seeking more reliable evidence from auxiliary sources, such as relevant news articles. This approach aligns with the principles of the Elaboration Likelihood Model (ELM) \cite{Petty1986} from the field of persuasion, which illustrates the intuitive and logical paths in human information processing. To this end, we propose the \textbf{E}vidence-aware \textbf{M}ulti-source \textbf{I}nformation \textbf{F}usion (EMIF) network, a novel framework designed to effectively collect objective evidence and enhance the robustness of interpretable fake news detection model. We achieve this by implementing co-attention mechanisms that capture global semantic conflicts between news content and user comments. In the meantime, we compare the original news with relevant news and select the representative top-K articles with consistent topics but the largest semantic divergence. To ensure consistency in evidence collection between user comments and relevant news while reducing individual cognitive bias, we introduce an ‘inconsistent loss’ function to penalize divergence. Experimental results reveal the outstanding detection performance and also indicate remarkable explainability of the proposed EMIF.

\section{Related Work}

Popular fact-checking sites like Snopes.com, PolitiFact.com, and FactCheck.org rely on manual operation approaches, including expert reviews and crowdsourcing techniques. However, these conventional methods become increasingly time-consuming and labor-intensive as the volume of messages grows. Consequently, various research has been conducted on automatic detection of fake news \cite{Zhang2020}. The existing work can be categorized into two main aspects: models for fake news detection and the sources of information as input in fake news detection.

\subsection{Models for fake news detection}

The evolution of fake news detection has seen three key stages. (i) \textbf{Traditional machine learning based detection model}. Early studies on fake news detection employ multiple machine learning methods (e.g., Random Forest \cite{Kwon2013}, SVM \cite{Yang2012} \cite{Ma2015}, Logistic Regression \cite{Bondielli2019}, Bayes \cite{Qazvinian2011}, etc.). These models are well-performed in terms of small datasets. However, due to their heavy reliance on hand-crafted feature engineering \cite{Horne2017} \cite{Wang2017}, the mentioned approaches tend to be highly labor-consuming and easily subjective to bias \cite{Liu2016}. (ii) \textbf{Deep learning based detection model}. Deep learning algorithms (e.g., CNN \cite{Yu2017}, GAN \cite{Ma2019}, BiLSTM \cite{Bahad2019}, hybrid models \cite{Nasir2021} \cite{Liu2018}, etc.) have excelled in capturing semantic \cite{Wu2021c}, emotional \cite{Zhang2021} \cite{Xue2022}, stance-based \cite{Hardalov2021} and stylistic \cite{Sheng2021} \cite{Zeng2022} features from raw data \cite{LeCun2015}. However, these neural-networks-based methods provide little insight into how results are derived owing to their black-box attributions. (iii) \textbf{Interpretable detection model}. Aiming at providing human operators with interpretable AI models and understandable AI decisions, Explainable AI (XAI) \cite{Arrieta2020}, has aroused increasing attention in recent years. One prevalent approach in XAI for complex deep learning models is to utilize rule extraction techniques, such as LIME \cite{Ribeiro2016}, which can derive a simplified model reflecting the working mechanism of the original complex model. However, this approach may not be universally feasible and could yield explanations unsuitable for various users \cite{Alharbi2021}. In contrast, feature relevance methods \cite{Ayoub2021} generate intrinsic explanations by assigning relevance scores to input variables, quantifying their contributions to model predictions. In the realm of fake news detection, the fundamental principle is to quantify the association and interaction between news content and corresponding comment features (or external knowledge), serving as evidence to expose falsehoods within fake news \cite{Nie2019}. This improvement does bolster the understandability and reliability of the detection model \cite{Mohseni2019}. Moreover, researchers have explored attention mechanisms, such as co-attention networks, to jointly analyze posts and comments and capture relevant evidence sentences for explainable fake news detection\cite{Shu2019a} \cite{Khoo2020}. Building upon this foundation, our work is centered on developing an explainable fake news detection network with an attention mechanism.

\subsection{Information sources in fake news detection}

Previous research in fake news detection falls into two categories based on input features: social-context-based and content-based methods \cite{Wang2023}. Social-context-based methods exploit the overall social activity system in which the news disseminated, including the distribution of social data \cite{Zhao2014}, user characteristics \cite{Shu2019} and their interaction networks \cite{Shu2019b}. However, capturing social context features can be resource-intensive, leading recent approaches in online fake news detection to primarily focus on direct content analysis \cite{Zhang2020}. Except for using \textbf{original fake news content} to capture the discriminative features, such as linguistic patterns and writing styles \cite{Castillo2011} \cite{Wu2021c}, from truth news, researchers have utilized auxiliary evidence or knowledge for news verification. For instance, comments, as a common information source, have been widely used as robust evidence to enhance detection performance \cite{Wu2023} and interpretability \cite{Shu2019a} \cite{Ge2022}. Meanwhile, \textbf{relevant news/claims} also contribute as pieces of evidence in the task of news/claims verification. A series of interactive models construct correlations between news and relevant news/claims to explore conflicting \cite{Wu2021b}, coherent \cite{Wu2021} or similar \cite{Nie2019} semantics as evidence for detecting the falsehood within news. However, in fake news detection, the quality of user comments may be regularly interfered by emotional bias \cite{Kumari2022}, exposure bias \cite{Greenwald1998}, cognitive bias \cite{Wu2021} and global noise \cite{Sun2022} (including unintentional misspelling and intentional camouflage strategies, such as deleting opposed comments, adding fake supportive comments, and account-hijacking). Conversely, relying solely on relevant news as a source of evidence presents concerns like data scarcity, data damage, and data obsolescence \cite{Wu2021a}. To address these challenges, we adopt insights from certain research in which \textbf{multi-source information} fusion strategies are employed \cite{Zhang2023} \cite{Xie2023}, jointly leveraging user comments and relevant news to enhance the robustness of the fake news detection and provide explainable prediction results with higher objectivity and credibility.

\section{Methodology}

\subsection{Background \& notations}

Formally, consider $A=\left\{a_{1}, a_{2, \ldots,} a_{N}\right\}$ as a news article containing $N$ sentences, with each sentence $a_{i}=\left\{w_{1}^{i}, w_{2}^{i}, \ldots, w_{n}^{i}\right\}$ containing $n$ words. We assume that news $A$ generates a set of comments $C=\left\{c_{1}, c_{2, \ldots,} c_{M}\right\}$, where each comment $c_{j}=\left\{w_{1}^{j}, w_{2}^{j}, \ldots, w_{m}^{j}\right\}$ containing $m$ words. $\left\{A_{r}^{\prime}\right\}_{r=1}^{R}$ is the set of all relevant news with quantity $R$ from several sources, and $A_{r}^{\prime}=\left\{A_{1}^{\prime r}, A_{2}^{\prime r}, \ldots, A_{l}^{\prime r}\right\}$ indicates the $r^{th}$ relevant news containing $l$ words.

Given a news $A$, along with the corresponding comments $C$ and a set of  relevant news $\left\{A_{r}^{\prime}\right\}_{r=1}^{R}$, we aim to predict the truthfulness $y$ of the news $A$. We approach the fake news detection task as a binary classification problem, framing it within the context of a binary label $y \in\{0,1\}$ representing the truthfulness. Specifically, we define $y=0$ to indicate the veracity of the news and $y=1$ to signify its falsity. Additionally, we demand that our algorithm automatically identify the semantic inconsistencies within the source news and select the top $K$ relevant news that best elucidate why $A$ is determined to be either true or false.

\subsection{Model Architecture}

\begin{figure}[h!] 
	\centering 
	\includegraphics[width = 6in]{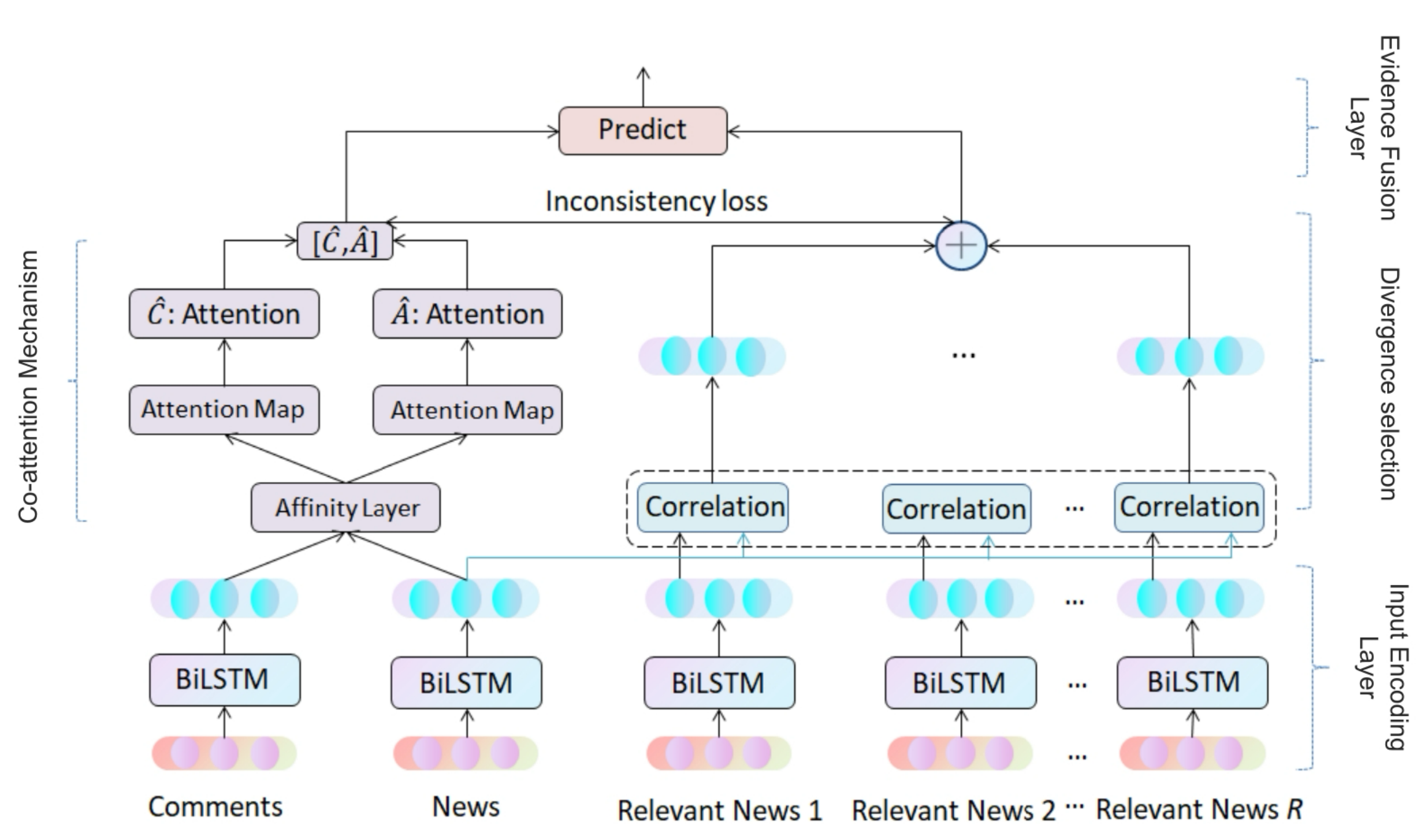} 
	\caption{The architecture of EMIF model.} 
	\label{f2}
\end{figure}

We describe the overall architecture of the proposed EMIF in detail here. In our approach, we prioritize user comments as primary source of information for fake news detection, with relevant news serving as an auxiliary information source. Notably, the inconsistency between comments and relevant news is an issue that must be considered in multi-source information fusion. Hence, in our proposed method, an inconsistency loss is adopted to penalize the disagreement between these two evidence sources during the fusion process. As shown in Fig. \ref{f2}, EMIF comprises four key components: (i) input encoding layer, (ii) co-attention mechanism, (iii) divergence selection, and (iv) evidence fusion layer.

\subsection{Input encoding layer}
EMIF has three types of inputs: news to be verified, corresponding user comments and relevant news. Since bidirectional long and short-term memory (BiLSTM) \cite{Ashik2021} both maintains a more persistent memory and captures contextual information about the annotations, we utilize it to encode words in both directions. Particularly, we use an embedding matrix to transform each word $w_{t}^{i}, t \in\{1, \ldots, n\}$ in a given sentence $a_{i}$ into its corresponding word vector $\mathbf{w}_{t}^{i} \in \mathbb{R}^{d}$. Subsequently, the feedforward and backward hidden states $\overrightarrow{\mathbf{h}_{t}^{i}}$ and  $\overleftarrow{\mathbf{h}_{t}^{i}}$ are obtained:

\begin{align}
&\overrightarrow{\mathbf{h}_{t}^{i}}=\overrightarrow{\operatorname{LSTM}}\left(\mathbf{w}_{t}^{i}\right), t \in\{1, \ldots, n\} \\
&\overleftarrow{\mathbf{h}_{t}^{i}}=\overleftarrow{\operatorname{LSTM}}\left(\mathbf{w}_{t}^{i}\right), t \in\{n, \ldots, 1\}
\end{align}

After concatenating $\overrightarrow{\mathbf{h}_{t}^{i}}$ and $\overleftarrow{\mathbf{h}_{t}^{i}}$, we obtain a comprehensive sentence annotation $\mathbf{h}_{t}^{i}=\left[\overrightarrow{\mathbf{h}_{t}^{i}},  \overleftarrow{\mathbf{h}_{t}^{i}}\right]$ which captures the entire content of the sentence. Since each word plays a different role in the news article, they should be assigned different attention. Therefore, an attention mechanism is utilized to give varied weights to words of varying importance in a news article. The sentence vector $\mathbf{a}_{i}=\left\{\mathbf{h}_{1}^{i}, \ldots, \mathbf{h}_{n}^{i}\right\} \in \mathbb{R}^{2 d}$ is computed as follows:

\begin{equation}
\mathbf{a}_{i}=\sum_{t=1}^{n} \alpha_{t}^{i} \mathbf{h}_{t}^{i}
\end{equation}

\noindent
where $\alpha_{t}^{i}$ denotes how significant the $t^{th}$ word in the $i^{th}$ sentence is, and we can calculate $\alpha_{t}^{i}$ as follows:

\begin{equation}
\alpha_{t}^{i}=\frac{\exp \left(\mathbf{u}_{t}^{i} \mathbf{u}_{w}^{\top}\right)}{\sum_{e=1}^{n} \exp \left(\mathbf{u}_{e}^{i} \mathbf{u}_{w}^{\top}\right)}
\end{equation}

\begin{equation}
\mathbf{u}_{t}^{i}=\tanh \left(\mathbf{W}_{w} \mathbf{h}_{t}^{i}+\mathbf{b}_{w}\right)
\end{equation}

\noindent
where $\mathbf{u}_{w}$ is a weight parameter representing the context vector. When a fully-embedding layer is fed the hidden state $\mathbf{h}_{t}^{i}$, it produces $\mathbf{u}_{t}^{i}$.

Finally, we obtain the whole news article’s representation as $\mathbf{A}=\left[\mathbf{a}_{1}, \mathbf{a}_{2}, \ldots, \mathbf{a}_{N}\right] \in \mathbb{R}^{2 d \times N}$ , which can also be represented in the form of word-level representations as $\mathbf{A}=\left\{\mathbf{h}_{1}, \mathbf{h}_{2}, \ldots, \mathbf{h}_{N \times n}\right\} \in \mathbb{R}^{2 d}$. In analogy to the procedure for news encoding, we utilize BiLSTM to model the word sequences within user comments and relevant news, with the attention mechanism being applied to learn the weights. Subsequently, each comment vector $\mathbf{c}_{j} \in \mathbb{R}^{2 d}$ and each relevant news $\mathbf{A}_{r}^{\prime}=\left\{\mathbf{h}_{1}, \mathbf{h}_{2}, \ldots, \mathbf{h}_{l}\right\} \in \mathbb{R}^{2 d}$ vector can be obtained along these lines.

\subsection{Co-attention mechanism}

It is crucial to acknowledge that the presence of fake news does not necessarily entail the fabrication of every sentence. Similarly, not every comment directly challenges the erroneous aspects of the news, as individuals may focus on different facets of the content. Some comments may address alternative controversial viewpoints, while others may simply introduce noise into the discussion. Our goal is to unveil the evidence for identifying fake news through investigating the specific parts of the news content that are addressed in particular user comments. Thus, our model employs a co-attention mechanism, a widely adopted technique in such detection tasks, to capture global dependencies across all positions in a sequence \cite{Lu2016}. With co-attention learning, our model gains interpretability by examining the attention weights between news sentences and comments concurrently. Specifically, given a news feature matrix  $\mathbf{A}=\left[\mathbf{a}_{1}, \mathbf{a}_{2}, \ldots, \mathbf{a}_{N}\right] \in \mathbb{R}^{2 d \times N}$ and a feature matrix of comments set $\mathbf{C}=\left[\mathbf{c}_{1}, \mathbf{c}_{2}, \ldots, \mathbf{c}_{M}\right] \in \mathbb{R}^{2 d \times M}$, we first compute the affinity matrix $\mathbf{F} \in \mathbb{R}^{M \times N}$ as follows:

\begin{equation}
\mathbf{F}=\tanh \left(\mathbf{C}^{\top} \mathbf{W}_{l} \mathbf{A}\right)
\end{equation}

\noindent
where $\mathbf{W}_{l} \in \mathbb{R}^{2 d \times 2 d}$ contains learnable weights. 

Instead of implementing the max activation, we adopted the suggestion in \cite{Lu2016} to treat the affinity matrix as a feature. Following that, we can train the model to predict attention maps for both news sentence and comments, given by

\begin{align}
\mathbf{H}^{a}=\tanh \left(\mathbf{W}_{a} \mathbf{A}+\left(\mathbf{W}_{c} \mathbf{C}\right) \mathbf{F}\right) \\
\mathbf{H}^{c}=\tanh \left(\mathbf{W}_{c} \mathbf{C}+\left(\mathbf{W}_{a} \mathbf{A}\right) \mathbf{F}^{\top}\right)
\end{align}

\noindent
where $\mathbf{W}_{a}, \mathbf{W}_{c} \in \mathbb{R}^{k \times 2 d}$ are matrices of learnable parameters. The affinity matrix  $\mathbf{F}$ can be thought to transform comments attention space to news attention space (vice versa for $\mathbf{F}^{\top}$).

The following is how we calculate attention values $\boldsymbol{v}^{a} \in \mathbb{R}^{1 \times N}$ for each sentence $a_{i}$ of the news and $\boldsymbol{v}^{c} \in \mathbb{R}^{1 \times M}$ for each user comment $c_{j}$,

\begin{align}
\boldsymbol{v}^{a}=\operatorname{softmax}\left(\mathbf{w}_{h a}^{\top} \mathbf{H}^{a}\right) \\
\boldsymbol{v}^{c}=\operatorname{softmax}\left(\mathbf{w}_{h c}^{\top} \mathbf{H}^{c}\right)
\end{align}

\noindent
where $\mathbf{w}_{h a}$,$\mathbf{w}_{h c}$ represent attention probabilities of each sentence in the original news and each piece of comment, respectively. The attention vectors of news sentences and comments can be generated through a weighted sum using the above attention weights, i.e.,

\begin{align}
&\widehat{\mathbf{A}}=\sum_{i=1}^{N} \boldsymbol{v}_{i}^{a} \mathbf{a}_{i} \\
&\widehat{\mathbf{C}}=\sum_{j=1}^{M} \boldsymbol{v}_{j}^{c} \mathbf{c}_{j}
\end{align}

\noindent
where $\widehat{\mathbf{A}} \in \mathbb{R}^{2 d}$ and $\widehat{\mathbf{C}} \in \mathbb{R}^{2 d}$ are the learned co-attention feature vectors for news sentences and user comments.

Eventually, we further integrate weighted feature representation of original news $\widehat{\mathbf{A}}$ and user comments $\widehat{\mathbf{C}}$ by concatenation operation, so that we obtain a representation $[\widehat{\mathbf{A}}, \widehat{\mathbf{C}}]$ capturing both context information of the news and semantic conflicts.

\subsection{Divergence selection}

In contrast to user comments, relevant news from diverse sources converges multiple perspectives and thus facilitates a more objective and comprehensive depiction of the truth. When the original news and its corresponding relevant news present conflicting viewpoints on the same topic, a significant divergence emerges in their descriptions, reflected in substantial differences in their vector representations. Following this routine, we calculate the divergence in vector representations and select the top-K representative relevant news with the largest semantic divergence. This process lays the groundwork for assessing the authenticity of the news.

To do this, the selected mechanism learns a vector $\mathbf{S} \in \mathbb{R}^{1 \times R}$ to restore the similarity values between original news and each relevant news in an automated manner. The entry of $\mathbf{S}$ is computed as follows:

\begin{align}
&\mathbf{u}=\varphi(\mathbf{W} \mathbf{A}+\mathbf{b}) \\
&\mathbf{u}_{k}=\varphi\left(\mathbf{W}_{r} \mathbf{A}_{r}^{\prime}+\mathbf{b}_{r}\right)\\
&\mathbf{S}[r]=\frac{\exp \left(\mathbf{u} \odot \mathbf{u}_{r}\right)}{\sum_{e=1}^{R} \exp \left(\mathbf{u}_{e} \odot \mathbf{u}_{r}\right)}
\end{align}

\noindent
where $\mathbf{W}$ and $\mathbf{W}_{r}$ are learnable weight matrix, $\mathbf{b}$ and $\mathbf{b}_{r}$ are biases, $\odot$ stands for dot product operator, and $\varphi$ denotes an activation function. A larger $\mathbf{S}[r]$ symbolizes the higher similarity between the original news and $r^{th}$ relevant news. Correspondingly, a smaller $\mathbf{S}[r]$ represents a greater semantic conflict between the original news and the relevant news. In the end, we filter the top-K relevant news with high divergence and integrate them through concatenation operation.

\begin{equation}
\mathbf{A}^{\prime}=\left[\mathbf{A}_{1}^{\prime}, \mathbf{A}_{2}^{\prime}, \ldots, \mathbf{A}_{K}^{\prime}\right]
\end{equation}

\subsection{Evidence fusion layer}

To guarantee the acquisition of decent performance of fake news detection, we put forward an evidence fusion strategy to refute the original news from both general and concrete perspectives. Initially, we introduce an inconsistency loss $ \mathcal{L}_{\mathrm{KL}} $ to enhance the consistency of evidence collection between user comments and relevant news, while simultaneously alleviating individual cognitive bias in comments. The inconsistency loss function is defined by Kulllback-Leibler (KL) divergence between $\widehat{\mathbf{A}^{\prime}}$ and $[\widehat{\mathbf{A}}, \widehat{\mathbf{C}}]$, compelling the two pieces of evidence to align as closely as possible during the screening process.

\begin{equation}
\mathcal{L}_{\mathrm{KL}}=\mathrm{D}_{\mathrm{KL}}\left(\mathbf{A}^{\prime} \|[\widehat{\mathbf{A}}, \widehat{\mathbf{C}}]\right)=\sum_{q=1}^{Q} \mathbf{A}_{q}^{\prime} \log \frac{\mathbf{A}_{q}^{\prime}}{[\widehat{\mathbf{A}}, \widehat{\mathbf{C}}]_{q}}
\end{equation}

\noindent
where $\mathbf{A}_{q}^{\prime}$ is the $q^{th}$ element of $\mathbf{A}^{\prime}$ and $[\widehat{\mathbf{A}}, \widehat{\mathbf{C}}]_{q}$ is the $q^{th}$ element in$[\widehat{\mathbf{A}}, \widehat{\mathbf{C}}]$.

Additionally, $\mathcal{L}_{\mathrm{CE}}$ minimizes the cross-entropy loss of the news classification task, where a softmax function is used to generate the prediction of probability distribution for training:

\begin{align}
&\mathcal{L}_{\mathrm{CE}}=-\sum y \log p \\
&p=\operatorname{softmax}\left(\mathbf{W}_{p}\left[[\widehat{\mathbf{A}}, \widehat{\mathbf{C}}] ; \mathbf{A}^{\prime}\right]+\mathbf{b}_{p}\right)
\end{align}

\noindent
where $\mathbf{W}_{p}$ and $\mathbf{b}_{p}$ are the learnable parameters.

We combine the two losses for joint training to improve the training effectiveness and ensure the mutual restraint between the two evidence as well.

\begin{equation}
\mathcal{L}=\beta \mathcal{L}_1+\mathcal{L}_2
\end{equation}

\noindent
where $\beta$ is the hyperparameter.

\section{Experiments}

\subsection{Datasets}
A publicly available dataset FibVID \cite{Kim2021} is utilized for our approach evaluation. The dataset contains news with indicators of truth or false (T/F) which have been confirmed by Politifact and Snopes, corresponding user comments and text similarity information. To control the consistency of the news topic, only COVID-19 news was selected to utilize. In addition, we extended FibVID by incorporating additional verified news and their corresponding comments. This augmentation results in a more balanced sample of positive and negative instances in our datasets. Comprehensive statistics detailing the extended datasets are provided in Tab. \ref{t1}.

\begin{table}[h!]
	\centering
	\setlength{\abovecaptionskip}{0pt}     
	\setlength{\belowcaptionskip}{0.2cm}
	\caption{Statistics of the datasets}
	\setlength{\tabcolsep}{3mm}{
	\label{t1}
\begin{tabular}{ll} 
	\toprule
	Dataset&Number \\
	\midrule
	Total data&151162 \\
	\quad FibVID(COVID-19)&140716 \\
	\quad Supplements&10446 \\
	\midrule
	True news&1093 \\
	Fake news&1136 \\
	\midrule
	User Comments&148933 \\
	\midrule
	Avg.Relevant News per News&26.16 \\
	\bottomrule
\end{tabular}}
\end{table}

\subsection{Baselines}

We compare EMIF with several state-of-the-art baselines for fake news detection:

\begin{itemize}
	
	\item \textbf{SVM} \cite{Cusmaliuc2018}: The SVM classifier utilizes features extracted manually from relevant articles as input and generates an optimal hyperplane which categorizes the test data as true or false.
	
	\item \textbf{Text-CNN} \cite{Lai2015}: Text-CNN encodes news content through a convolutional neural network, capturing text features at different levels of granularity.
	 
	\item \textbf{StA-HiTPLAN} \cite{Khoo2020}: StA-HiTPLAN proposes a hierarchical attention model to learn sentence representation within each tweet at the token level and the post level.
	
	\item \textbf{HAN} \cite{Yang2016}: HAN captures word-level and sentence-level evidence by constructing a hierarchical attention network to analyze the interaction between claims and related articles, thereby considering thematic coherence and semantic inference strength.
	
	\item \textbf{dEFEND} \cite{Shu2019a}: dEFEND develops a news-comments interactive co-attention network to identify the top-K relevant corpora, which serve as evidence for fake news detection.
	 
\end{itemize}

\subsection {Overall performance}
Tab. \ref{t3} presents a comparison of the performance of EMIF against baseline models. We observe the following issues: (i) With traditional machine-learning-based method SVM being the weakest performer across all baselines, neural network-based methods such as Text-CNN outperform it by at least 7.7\% in terms of Accuracy. This outcome demonstrates the significant advantages of neural network models in feature extraction. (ii) On the basis of extraction of semantic features from news content, Text-CNN and StA-HiTPLAN achieve 61\% and 74.2\% in terms of Accuracy. With the additive enhancement of the hierarchical attention network, StA-HiTPLAN realize higher effectiveness in distinguishing true and false news. (iii) Additionally, HAN and dEFEND excavate semantics from the interaction between news content and user comments, showing up 4.6\% and 6.1\% Accuracy improvement over StA-HiTPLAN, respectively. (iv) Ultimately, our proposed EMIF remarkably outperforms the strongest baseline (dEFEND) by 3.5\% in Accuracy and 4.4\% in F1. The experimental results underscore the superiority of EMIF which integrates relevant news as a supplementary evidence source and employs inconsistency loss to filter out redundant noise.

\begin{table}[h!]
	\centering
	\setlength{\abovecaptionskip}{0pt}     
	\setlength{\belowcaptionskip}{0.2cm}
	\caption{The performance comparison of EMIF against the baselines}
	\setlength{\tabcolsep}{3mm}{
	\label{t3}
\begin{tabular}{lllll}
	\toprule
	Method&Accuracy&Precision&Recall&F1 \\
	\midrule
	SVM & $0.533$ & $0.533$ & $0.533$ & $0.533$ \\
	Text-CNN & $0.610$ & $0.623$ & $0.571$ & $0.578$ \\
	StA-HiTPLAN & $0.742$ & $0.705$ & $0.795$ & $0.747$ \\
	HAN & $0.788$ & $0.779$ & $0.815$ & $0.797$ \\
	dEFEND & $0.803$ & $0.767$ & $0.843$ & $0.803$ \\
	$\mathbf{EMIF}$ & $\mathbf{0 . 8 3 8}$ & $\mathbf{0 . 7 9 1}$ & $\mathbf{0 . 9 1 2}$ & $\mathbf{0 . 8 4 7}$ \\
	\bottomrule
\end{tabular}}
\end{table}

\subsection{Ablation study}
\begin{figure}[h!]
	\centering 
	\includegraphics[width = 4in]{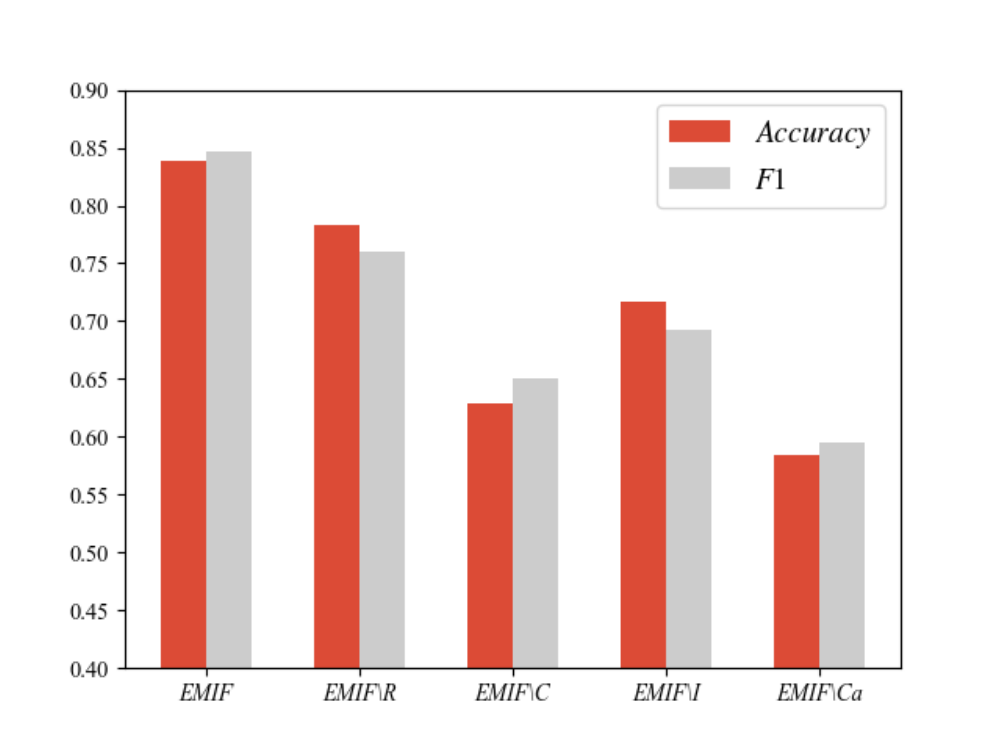} 
	\caption{Impact analysis of each component of EMIF for fake news detection} 
	\label{f3}
\end{figure}

We conduct a series of ablation experiments to evaluate the contributions of each key component in EMIF. Four simplified variants of EMIF are defined by stripping certain components away from the entire model. Specifically, “\textbf{$\backslash$R}” and “\textbf{$\backslash$C}” denote the variants which exclude information from relevant news and user comments, respectively. By removing co-attention mechanism, we have a variant “\textbf{$\backslash$CA}”. Besides, we define “\textbf{$\backslash$IL}” as the variant of EMIF without calculating inconsistency loss between user comments and relevant news, which indicates the separate selection of the two types of evidence. The performance of these variants is reported in Tab. \ref{t4} and Fig. \ref{f3}, yielding the following observations:

\begin{table}[h!]
	\centering
	\setlength{\abovecaptionskip}{0pt}     
	\setlength{\belowcaptionskip}{0.2cm}
	\caption{Ablation analysis of EMIF}
	\setlength{\tabcolsep}{3mm}{
	\label{t4}
\begin{tabular}{lll}
	\toprule
	Model&Accuracy&F1 \\
	\midrule
	$\mathbf{EMIF}$ & $\mathbf{83.80\%}$ & $\mathbf{84.73\%}$ \\
	EMIF$\backslash$R & $ 78.32\% $ & $76.01\% $ \\
	EMIF$\backslash$C & $ 62.89\% $ & $65.11\% $ \\
	EMIF$\backslash$I & $ 71.73\% $ & $69.19\% $ \\
	EMIF$\backslash$Ca & $ 58.46\% $ & $59.52\% $ \\
	\bottomrule
\end{tabular}}
\end{table}

\begin{itemize}
	
	\item Compared to EMIF, we find out Accuracy and F1 score of EMIF$\backslash$R are reduced by 5.48\% and 8.72\%, respectively. This result highlights the substantial effectiveness of relevant news as an evidence source, which aids in pinpointing core errors from massive interactive information.
	
	\item For EMIF$\backslash$C, the removal of the news-comments co-attention module severely weakens the performance of EMIF, resulting in a notable 20.91\% reduction in Accuracy and a 19.63\% reduction in F1 score. This decrease in performance was even more severe than that observed in EMIF$\backslash$R, emphasizing the dominant contribution of user comments in fake news detection.
	
	\item EMIF$\backslash$CA performs the poorest among all variants, with reduction of 25.34\% and 25.21\% in Accuracy and F1, respectively. When omitting co-attention mechanism from EMIF, some unrelated comments appearing as noise fail to be excluded, which further affects the filtering of relevant news through inconsistency loss. As a result, inputs of the evidence fusion layer are teeming with extraneous information, leading to inaccurate prediction.
	
	\item Finally, in EMIF$\backslash$IL, Accuracy dropped by 12.07\% and F1 decreased by 15.54\%. It suggests the necessity of incorporating the inconsistency loss function as a mutual constraint on the selection of user comments and relevant news.
	
\end{itemize}

\subsection{Explainability analysis}

In this subsection, we present a visualization of the output features learned from EMIF. As depicted in Fig. \ref{f4}, words with different attention weights are highlighted with different shades of color. It helps us understand what EMIF prioritizes and how these priorities influence its decisions, making our model transparent to the end-users. Our observations are as follows:

\begin{figure}[h!]	
	\centering 
	\includegraphics[width = 6in]{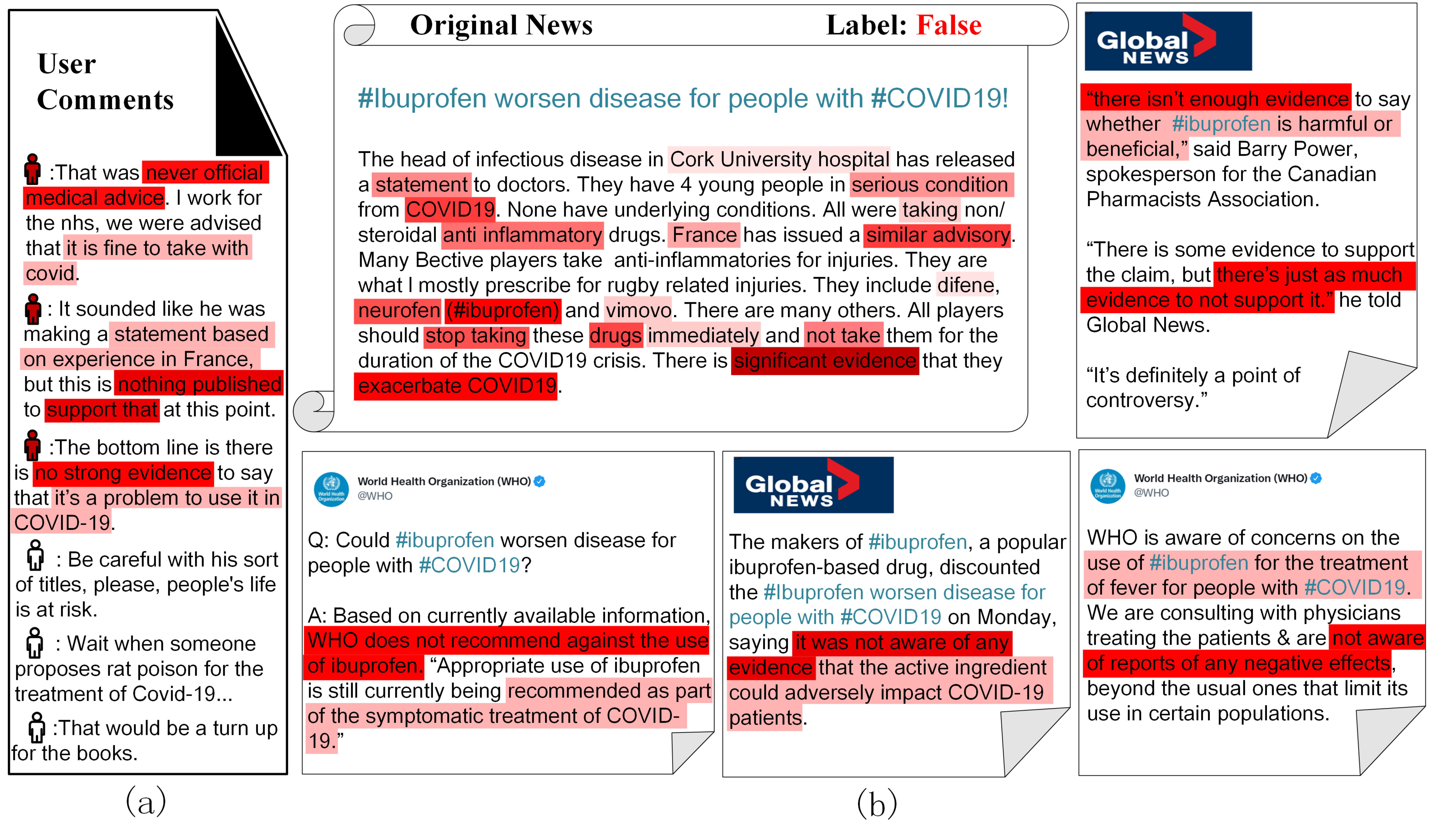} 
	\caption{Explainability via visualization of attention weights in EMIF. The label [True/False] indicates the verdict of original news. (a) are user comments (the first 3 are explainable comments captured can be used as evidence), (b) are original news (labeled false, darker shades correspond to higher-weighted words) and selected relevant news (only 4 items were presented).} 
	\label{f4}
\end{figure}

Distinguished from the comments which only represent subjective attitudes, the captured explainable comments demonstrate greater relevance to the original news in terms of content, as well as better capability of hitting the heart of the matter. For instance, the first three user comments listed in Fig.4(a) concurrently emphasize the lack of solid evidence for the statements in the original news, using phrases such as “...no strong evidence...”, “...nothing published to support that...” and “...never official medical advice...”. These comments correspond to the phrase “significant evidence” in the original text, highlighted with the darkest shades. Thus, EMIF accurately identifies valuable user comments through the co-attention mechanism, which pinpoints semantic conflicts within the original news.

In the meantime, the selected relevant news challenges the authenticity of original news from two perspectives. The first one questions whether ibuprofen harms infected individuals and emphasizes the need for further scientific verification, while the other illustrates that consultations with experts are still ongoing. By giving objective and detailed explanations, these relevant news statements craft a highly convincing narrative of the truth behind the story. In other words, those relevant news selected by EMIF accurately seize crucial conflicts and offer concrete evidence.

Both user comments and relevant news can reveal the potential falsehood of the original text, i.e. “there isn’t enough evidence”, which indicates that the topic deviation has been effectively prevented through the inconsistent loss. By doing so, redundant noise is filtered out from user comments and related news simultaneously, enabling us to provide both general and detailed complementary evidence.

\section{Conclusion}
In this paper, a novel multi-source information fusion network, EMIF, is constructed to collect evidence with higher objectivity and credibility in the explainable fake news detection task. Motivated by real-life evidence-aware identification of fake news, our proposed model innovatively combines user comments and relevant news as inputs, addressing a gap in previous research. General and concrete evidence are extracted through a news-comment co-attention mechanism and a divergence selection module for relevant news, respectively. Subsequently, we employ an inconsistent loss as a penalty to further filter out the redundant noise among these evidence. Numerical experiments on publicly available datasets demonstrate the effectiveness of our explainable framework. Moreover, attributed to its splendid robustness, EMIF is fully capable of being extended to complicated situations where a particular source of information is unavailable. For future work, we seek to enhance our model by considering varying data exposure levels and incorporating more informative content modalities, such as images and videos. In addition, insights from other interdisciplinary research (e.g., social cognition and psychology) hold the potential to further improve our explainability.

\bibliographystyle{elsarticle-num}
\bibliography{EMIF}

\end{document}